\documentclass[aps,prb,twocolumn,superscriptaddress]{revtex4-1}
\usepackage{graphicx}
\usepackage{amsmath}
\usepackage{amssymb}
\usepackage{graphicx}
\usepackage{afterpage}
\usepackage{amsfonts}
\usepackage{amssymb}
\usepackage{graphicx}
\usepackage{braket}

\newcommand{\be}{\begin{equation}}
\newcommand{\ee}{\end{equation}}
\newcommand{\bea}{\begin{eqnarray}}
\newcommand{\eea}{\end{eqnarray}}
\newcommand{\Ea}{\ensuremath{{\cal E}_1}}
\newcommand{\Eb}{\ensuremath{{\cal E}_2}}
\newcommand{\Ec}{\ensuremath{{\cal E}_3}}

\newcommand{\Ed}{\ensuremath{{\cal E}_{1,\,2,\,3}}}
\newcommand{\Er}{\ensuremath{{\cal E}_{\rm r}}}
\newcommand{\Eo}{\ensuremath{{\cal E}_{1,\,2,\,3}}}
\newcommand{\polho}{\relbar}
\newcommand{\polve}{|}

\begin{document}

%\preprint{}

\title{Harvesting, coupling and control of single exciton coherences\\ in photonic waveguide antennas}



\author{Q.~Mermillod}
\email{quentin.mermillod@neel.cnrs.fr} \affiliation{Univ. Grenoble
Alpes, F-38000 Grenoble, France} \affiliation{CNRS, Institut
N\'{e}el, "Nanophysique et semiconducteurs" group, F-38000 Grenoble,
France}

\author{T.~Jakubczyk}
\affiliation{Univ. Grenoble Alpes, F-38000 Grenoble, France}
\affiliation{CNRS, Institut N\'{e}el, "Nanophysique et
semiconducteurs" group, F-38000 Grenoble, France}

\author{V.~Delmonte}
\affiliation{Univ. Grenoble Alpes, F-38000 Grenoble, France}
\affiliation{CNRS, Institut N\'{e}el, "Nanophysique et
semiconducteurs" group, F-38000 Grenoble, France}

\author{A.~Delga}
\affiliation{Univ. Grenoble Alpes, F-38000 Grenoble, France}
\affiliation{CEA, INAC-PHELIQS, "Nanophysique et semiconducteurs"
group, F-38000 Grenoble, France}

\author{E.~Peinke}
\affiliation{Univ. Grenoble Alpes, F-38000 Grenoble, France}
\affiliation{CEA, INAC-PHELIQS, "Nanophysique et semiconducteurs"
group, F-38000 Grenoble, France}

\author{J-M.~G\'{e}rard}
\affiliation{Univ. Grenoble Alpes, F-38000 Grenoble, France}
\affiliation{CEA, INAC-PHELIQS, "Nanophysique et semiconducteurs"
group, F-38000 Grenoble, France}

\author{J.~Claudon}
\affiliation{Univ. Grenoble Alpes, F-38000 Grenoble, France}
\affiliation{CEA, INAC-PHELIQS, "Nanophysique et semiconducteurs"
group, F-38000 Grenoble, France}

\author{J.~Kasprzak}
\email[]{jacek.kasprzak@neel.cnrs.fr}  \affiliation{Univ. Grenoble
Alpes, F-38000 Grenoble, France} \affiliation{CNRS, Institut
N\'{e}el, "Nanophysique et semiconducteurs" group, F-38000 Grenoble,
France}

%\pacs{}


\begin{abstract}
We perform coherent non-linear spectroscopy of individual excitons
strongly confined in single InAs quantum dots (QDs). The retrieval
of their intrinsically weak four-wave mixing (FWM) response is
enabled by a one-dimensional dielectric waveguide antenna. Compared
to a similar QD embedded in bulk media, the FWM detection
sensitivity is enhanced by up to four orders of magnitude, over a
broad operation bandwidth. Three-beam FWM is employed to investigate
coherence and population dynamics within individual QD transitions.
We retrieve their homogenous dephasing in a presence of spectral
wandering. Two-dimensional FWM reveals off-resonant F\"{o}rster
coupling between a pair of distinct QDs embedded in the antenna. We
also detect a higher order QD non-linearity (six-wave mixing) and
use it to coherently control the FWM transient. Waveguide antennas
enable to conceive multi-color coherent manipulation schemes of
individual emitters.
\end{abstract}

%\date{\today}

\maketitle

Semiconductor quantum dots (QDs) embedded in properly designed
photonic nanostructures hold great promises for future quantum
information technologies\,\cite{LodahlRMP15}. These few-level
mesoscopic emitters can serve as bright sources of non-classical
states of light\,\cite{ClaudonNatPhot10, DousseNature11,
SchulteNature15} or mediate strong optical non-linearities, at the
single-photon level\,\cite{KasprzakNMa10, JavadiNatComm15}. An
excitonic QD transition also constitutes a localized qubit that can
be coherently manipulated on sub-ps timescale with optical
pulses\,\cite{ZrennerN02, RamsayPRL10, Fras15}. Via coupling QD
excitons with solid-state spins, one can combine fast radiative
emission with a long-lived spin coherence. A spin-photon interface,
a prerequisite for a distributed quantum network, has been recently
demonstrated with such systems\,\cite{GaoNatPhot2015}. In this
dynamic context, an accurate assessment and control of the excitonic
coherence is required. Furthermore, harnessing inter-exciton
coupling mechanisms is of utmost importance for implementing
non-local coherent control schemes.

Coherent non-linear spectroscopy provides avant-garde techniques for
all-optical coherent manipulation and readout of single
emitters\,\cite{Fras15}, in particular by investigating four-wave
mixing (FWM). As depicted in Fig.\,\ref{fig:Fig2}, such polarization
is driven by three, resonant, short laser pulses with electric field
amplitudes $\Ea$, $\Eb$ and $\Ec$. In the lowest order, FWM is
proportional to $\mu^4\Ea^{\star}\Eb\Ec$, with $\mu$ denoting the
optical dipole moment of the emitter. By taking advantage of the
photon echo formation\,\cite{ChemlaNature01}, FWM has been exploited
in the past to infer spectral homogenous broadening in ensembles of
optical transitions in solids, even in a presence of inhomogeneous
broadening. Due to the steep dependence on $\mu$, first
investigations of FWM on \emph{single} emitters have been limited to
giant oscillator strength excitons confined by interface
fluctuations of a quantum well\,\cite{LangbeinPRL05,KasprzakNPho11}.
A successful strategy to improve the FWM retrieval relies on
photonic structures that locally enhance the driving fields
experienced by the QD and improve the collection efficiency of the
generated nonlinear response. In particular, semiconductor
micro-cavities have shown appealing prospects\,\cite{KasprzakNMa10,
AlbertNatComm13, Fras15}, albeit at the cost of an operation
bandwidth limited to the cavity resonance.

In this Letter, we show that waveguide antennas, initially
introduced to realize bright sources of quantum
light\,\cite{ClaudonNatPhot10,MunschPRL13}, dramatically enhance the
non-linear response of individual QDs over a broad spectral range.
This enhancement enables a comprehensive investigation of the
coherence properties of excitonic complexes strongly confined in
self-assembled InAs QDs. While such emitters constitute one of the
leading systems for solid-state quantum optics\,\cite{LodahlRMP15,
SchulteNature15}, their moderate oscillator strength has hindered
investigations of coherence at the single QD level. Furthermore, we
show that FWM can reveal an off-resonant coherent coupling via
underlying Coulomb interaction between two distinct QDs embedded in
the antenna. Finally, we recover six-wave mixing (SWM) of an
exciton-biexciton system, and use it to coherently control the FWM
transient. Our findings pave the way towards achieving non-local,
coherent control in small sets of individual emitters in a solid.

\begin{figure}[!ht]
\includegraphics[width=\columnwidth]{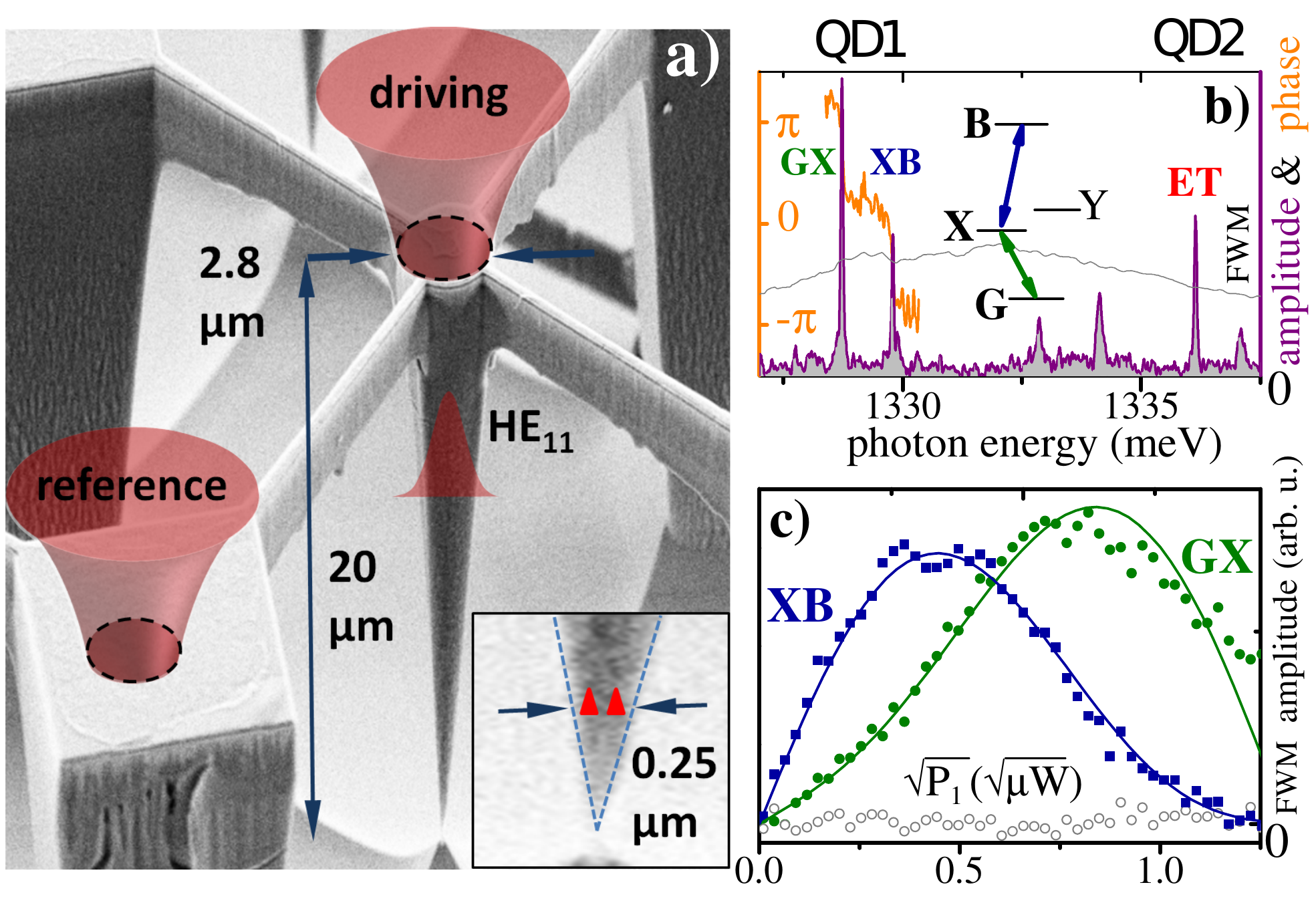}
\caption{\textbf{Nanowire antenna for enhancing the four-wave mixing
(FWM) response of excitonic transitions strongly confined in InAs
QDs.} a)\,SEM image of a suspended photonic trumpet, which is made
of GaAs and embeds a few self-assembled InAs QDs close to its base
(inset). This photonic structure simultaneously improves the
in-coupling of the resonant driving pulses to the QDs and the
out-coupling of the non-linear signal to the collection optics. Also
see section A in the SM. b)\,Spectrally-resolved FWM amplitude
(violet) and phase (orange). The pulse sequence of
Fig.\,\ref{fig:Fig2}\,a is employed, with $\tau_{12}=10\,$ps,
$(\theta_1,\,\theta_2)=(\pi/5,\,2\pi/5)$. The excitation spectrum is
given with a gray line. c)\,FWM amplitude as a function of $\Ea$
pulse area $\theta_1\propto\sqrt{{\rm P}_1}$, showing Rabi
oscillation at GX (green circles) and XB (blue squares) transitions.
The open symbols give the noise floor. The theoretical predictions
appear as solid lines. Measurement parameters: $\theta_2=0.82\pi$,
$\tau_{12}=5$\,ps. \label{fig:Fig1}}
\end{figure}

\begin{figure}[!ht]
\includegraphics[width=\columnwidth]{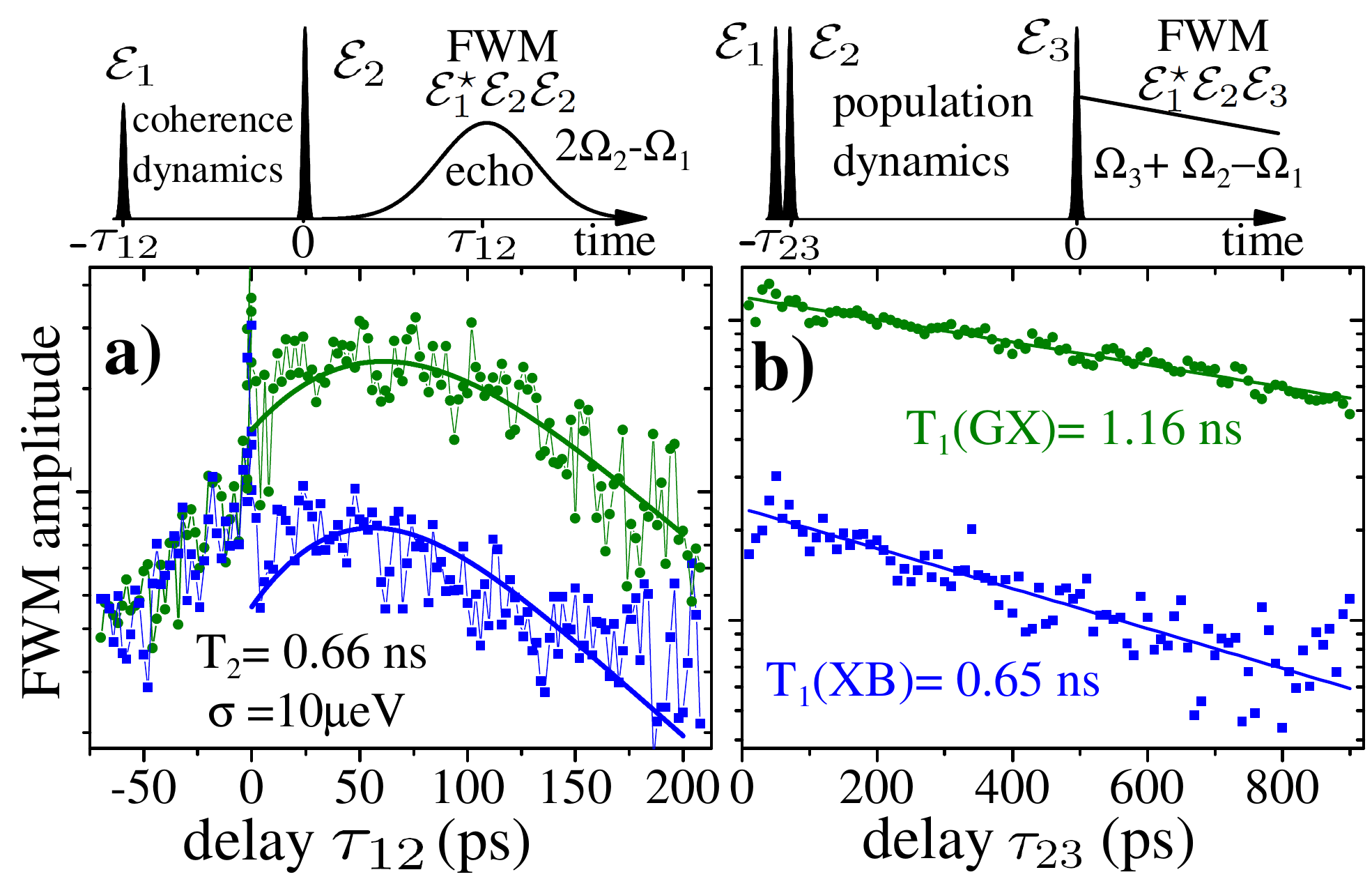}
\caption{\textbf{Coherence and population dynamics in the
exciton-biexciton system of QD1.} a)\,Coherence dynamics is
investigated via degenerate FWM. The top panel indicates the pulse
sequence and the related observable. The dependence of the FWM
signal on $\tau_{12}$ yields T$_2$ and $\sigma$, for both GX and XB
transitions. Applied pulse areas:
$(\theta_1,\theta_2)=(\pi/5,\,2\pi/5)$. b)\,Population dynamics is
investigated via non-degenerate FWM (top panel). The dependence of
the FWM signal on $\tau_{23}$ yields T$_1$ for X and B states.
Measurement parameters: $\tau_{12} = 1\,$ps,
$(\theta_1,\,\theta_2,\,\theta_3) = (\pi/2,\,\pi/2,\,\pi/2)$. Both
graphs feature logarithmic vertical scales. \label{fig:Fig2}}
\end{figure}

The antenna, shown in Fig.\,\ref{fig:Fig1}\,a, is a suspended GaAs
photonic trumpet\,\cite{MunschPRL13}\,(PT), which is anchored to
square pillars for improved mechanical stability. The driving pulses
$\Ed$ are focused on the circular top facet. Thanks to the nearly
Gaussian profile of the fundamental guided mode
(HE$_{11}$)\,\cite{StepanovAPL15}, Gaussian driving beams with
adapted waists are transmitted to the tapered section with
negligible losses. On the waveguide axis, the electrical field
amplitude scales as $\sqrt{n_g/S_\text{eff}}$, with $n_g$ the group
index and $S_\text{eff}$ the effective mode
surface\,\cite{BleusePRL11}. Field enhancement reaches a maximum at
the QDs position, when the waveguide diameter is reduced down to
$0.25\,\mu$m, mainly because of the optimal lateral confinement of
the mode, assisted by a modest slow-light
effect\,\cite{BleusePRL11}. Compared to a QD embedded in bulk GaAs,
a factor of $\sim 100$ is gained on the global amplitude
$\Ea\Eb\Ec$, which drives the FWM. Conversely, the collection
efficiency of the generated FWM increases from around $1\%$ up to
$45\%$. As a consequence, the FWM is retrieved with
signal-to-background ratio improved by a factor of $\sim4\times
10^3$. Moreover, these non-resonant photonic structures naturally
provide a large operation bandwidth ($> 100\:
$nm)\,\cite{ClaudonCPC13}.

The wave-mixing signals are retrieved by heterodyne spectral
interferometry technique\,\cite{LangbeinPRL05}. Employing
acousto-optic modulation, the driving pulses $\Ed$ are frequency
up-shifted by radio-frequencies $\Omega_{1,2,3}$ introducing
controlled phase-drifts in their respective trains. After having
acquired delays $\tau_{12}$ and $\tau_{23}$, $\Ed$ are recombined
into a common spatial mode and are focused on the sample with a
microscope objective. An unmodulated reference beam is focused on
the auxiliary pillar (see Fig.\,\ref{fig:Fig1}\,a) and the searched
response is discriminated in the reflectance by the phase-sensitive
optical heterodyning, attaining a selectivity in field (intensity)
of $10^6$ ($10^{12}$). The experimental setup is described in
Ref.\,[\onlinecite{Fras15}]. Measurements are conducted on a
structure similar to the one shown in Fig.\,\ref{fig:Fig1}\,a,
maintained at T$=5.2\pm0.5\,$K.

To explore the coherence dynamics of QD excitons, we first
concentrate on degenerate FWM. It is generated by a two-pulse
sequence, as depicted in Fig.\,\ref{fig:Fig2}\,a and retrieved at
the $2\Omega_2-\Omega_1$ heterodyne frequency. The
spectrally-resolved FWM amplitude and phase are shown in
Fig.\,\ref{fig:Fig1}\,b. The sharp peaks correspond to individual
excitonic transitions. Having verified the FWM polarization
selection rules, the resonances GX and XB are unequivocally
recognized as ground state\,-\,exciton and exciton\,-\,biexciton
transitions, hosted by a first QD, labeled QD1 [see section C of the
Supplementary Material (SM)]. Note the negative biexciton
renormalization energy of $\Delta\simeq-1\,$meV, indicating a small
spatial extent of wavefunctions of the both complexes and thus their
small $\mu$. Such system is a generic excitation in neutral QDs, in
particular enabling bright generation of entangled photon pairs from
solid state devices\,\cite{StevensonN06}. ET, which will be
discussed latter, is attributed to a charged exciton (trion), hosted
by a different QD, labeled QD2.

To directly illustrate the enhanced in-coupling of the driving
pulses, we present in Fig.\,\ref{fig:Fig1}\,c FWM amplitudes of GX
and XB as a function of $\Ea$ pulse area $\theta_1$ (being
proportional to the square-root of pulse $\Ea$ intensity P$_1$). The
FWM of both transitions reaches its maximum, corresponding to the
$\theta_1=\pi/2$, for impressively low $\Ea$ intensity of around
0$.3\,\mu$W, two orders of magnitude less than in strongly-confined
GaAs QDs\,\cite{KasprzakNJP13} embedded in bulk. Further increase of
$\theta_1$ results in Rabi flopping modeled using analytical
prediction\,\cite{KasprzakNJP13} only by adjusting XB to GX dipole
moment ratio $\mu_{\rm XB}/\mu_{\rm GX}=0.9$ (see solid lines in
Fig.\,\ref{fig:Fig1}\,c).

As a first application of enhanced FWM harvesting, we conduct a
comprehensive investigation of the coherence and population dynamics
of individual excitonic transitions hosted by a QD, a task that is
relevant for all applications of this system to quantum information
technologies. Coherence dynamics is investigated by measuring FWM as
a function of the pulse separation $\tau_{12}$, as depicted in
Fig.\,\ref{fig:Fig2}\,a. Importantly, this technique allows to
disentangle homogenous and inhomogeneous dephasing processes. The
former is due to radiative recombination and pure dephasing, while
the latter is associated with spectral fluctuations of the
transition energy\,\cite{KasprzakNJP13}. As a representative
example, we consider in Fig.\,\ref{fig:Fig2}\,a the transitions GX
and XB hosted by QD1. The initial rise for $\tau_{12}>0$, followed
by an exponential decay, is due to the formation of a photon echo
triggered by a spectral wandering, which acts as a source of
inhomogeneous broadening (other examples are given in section B of
the SM). Using the model presented in
Ref.\,[\onlinecite{KasprzakNJP13}], we obtain the dephasing time
T$_2=(0.66\,\pm\,0.05)\,$ns, in agreement with measurements
performed on ensembles\,\cite{BorriPRB05}. In addition, the data
reveal a Gaussian inhomogeneous broadening, characterized by a
full-width at half maximum $\sigma=(10\,\pm\,4)\,\mu$eV. Also, it is
worth to note the presence of the signal for $\tau_{12}<0$,
generated by a two-photon coherence in a four-level
exciton-biexciton system\,\cite{KasprzakNJP13}. From the FWM decay
at negative delays we estimate two-photon dephasing time of T$_{\rm
TPC}=(41\,\pm\,13)\,$ps.

Population evolution is investigated with a three pulse sequence,
shown in Fig.\,\ref{fig:Fig2}\,b. Here, $\Ea$ and $\Eb$ first create
X and B population oscillating at $\Omega_2-\Omega_1$. The last
pulse, $\Ec$, creates a polarization proportional to this density,
the FWM, which is detected at the $\Omega_3+\Omega_2-\Omega_1$
frequency. Thus varying delay $\tau_{23}$ between $\Eb$ and $\Ec$,
the measured FWM reflects the population evolution in a QD. From the
resulting exponential decays of the FWM amplitude we retrieve the
exciton and biexciton lifetimes of T$_1({\rm
GX})=(1.16\,\pm\,0.04)\,$ns and T$_1({\rm
XB})=(0.65\,\pm\,0.05)\,$ns, respectively, with their ratio
approaching theoretical limit\,\cite{BacherPRL99} of two.

These measurements show that the investigated transitions are not
radiatively limited. For GX, T$_2$ is $3.5$ times smaller than
2T$_1$ and this factor is reduced to 2 for XB, which presents a
faster population decay. In addition, a significant spectral
wandering $\sigma$ is present. Such high- and low-frequency noises
are likely to be mainly caused by fluctuations of charge traps
located close to the QD\,\cite{BerthelotNatPhys06,
KuhlmanNatPhys13}, either in its immediate environment or in the
wire sidewall. While further research is necessary to elucidate this
point, the application of a dc electric field on the
QDs\,\cite{GregersenOE10, SomaschiNatPhot16} could help reducing
$\sigma$, further improving the coherence retrieval.

\begin{figure}[!ht]
\includegraphics[width=\columnwidth]{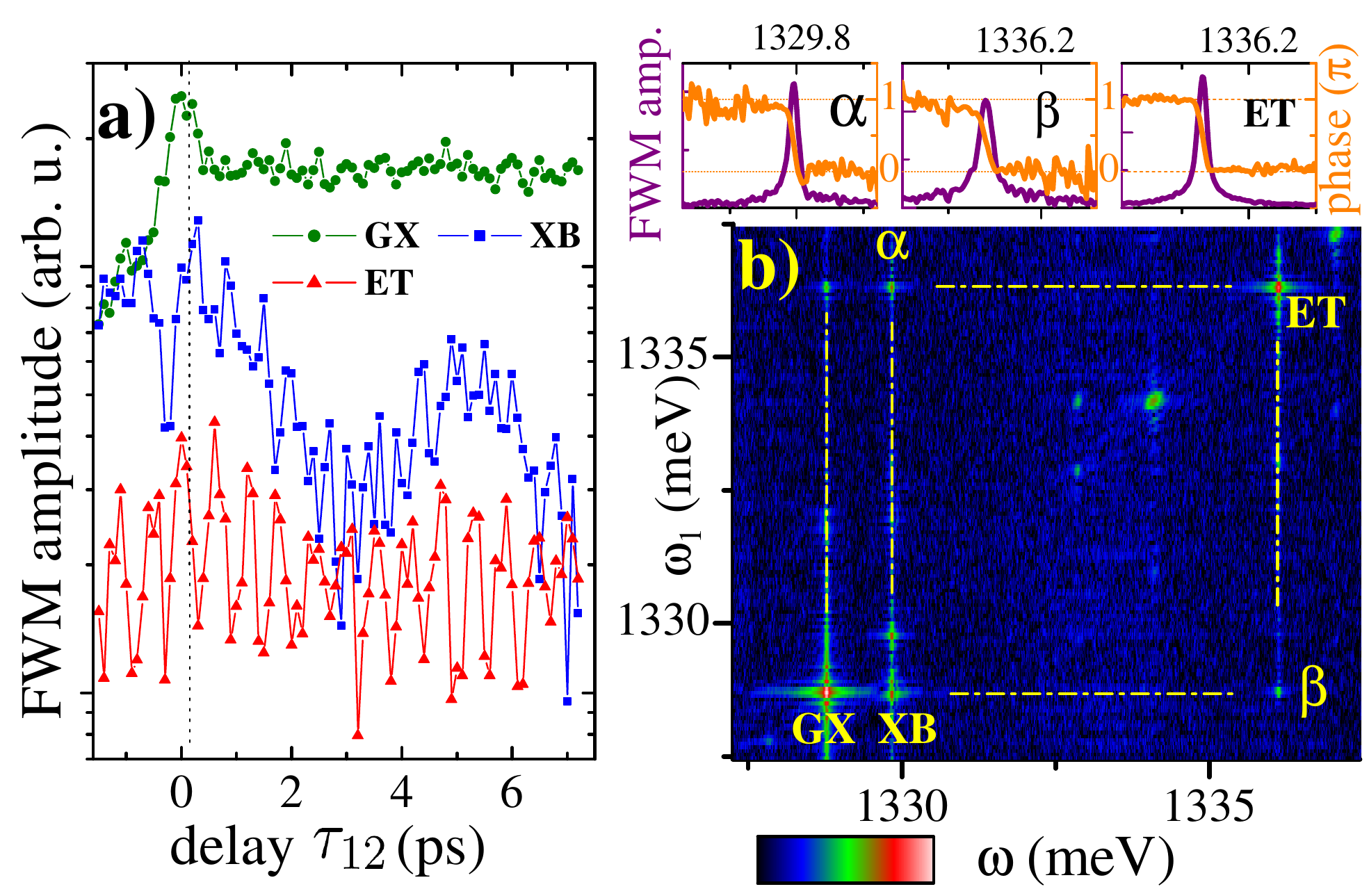}
\caption{\textbf{Coherent, off-resonant coupling between QD1 and
QD2}. We investigate here degenerate FWM (pulse sequence shown in
Fig.\,\ref{fig:Fig2}\,a) with a non-perturbative driving:
$(\theta_1,\,\theta_2)=(\pi/2,\,\pi)$. a)\,Delay dependence of the
FWM amplitude for GX, XB (QD1) and ET (QD2). b)\,Amplitude of the
two-dimensional FWM: coherent coupling is evidenced by a pair of
off-diagonal peaks, forming square features depicted with the dashed
lines. Note the presence of the diagonal peak for XB induced by the
non-perturbative driving. Logarithmic scale over two orders of
magnitude, given by the bar. Top:\,FWM amplitude and phase of the
off-diagonal peaks, labeled $\alpha$ and $\beta$ at $\omega=(1329.8,
1336.16)\,$meV and the diagonal ET peak.\label{fig:Fig3}}
\end{figure}

Beyond assessment of the single-QD coherence, we show below that FWM
can reveal a coherent\,\cite{KasprzakNPho11, AlbertNatComm13,
DanckwertsPRB06, SpechtPRB15}, non-resonant coupling between QD1 and
QD2. Establishing coherence transfer between a pair of emitters is
an essential ingredient to achieve optically controlled two-qubit
operations in solid. In a broader context, the coherence and charge
transfer is a multidisciplinary issue, spanning from biology, where
it is at heart of photosynthesis\,\cite{ScholesNCh11}, towards
quantum chemistry and photovoltaics. Here, we investigate FWM
simultaneously driven at GX, XB (QD1) and ET (QD2). As shown in
Fig.\,\ref{fig:Fig3}\,a, the coherence dynamics of XB is dominated
by the exciton-biexciton beating with a period
$|2\pi\hbar/\Delta|\simeq4.5\,$ps driven by fifth-order
contributions to the FWM\,\cite{TaharaPRB14}, predominant upon
$(\theta_1,\,\theta_2)\simeq(\pi/2,\,\pi)$ employed here.
Additionally, we observe a modulation with a period 0.58\,ps,
particularly pronounced on the ET transition and corresponding to
the spectral separation between both QDs, hence indicating their
mutual coupling.

Its definite display is provided by the two-dimensional FWM
spectrum\,\cite{KasprzakNPho11, AlbertNatComm13} shown in
Fig.\,\ref{fig:Fig3}\,b. By Fourier-transforming the FWM along the
delay $\tau_{12}$ we obtain a map linking the first-order absorption
frequency $\omega_1$, with the FWM one, $\omega$. The off-diagonal
signals in such a diagram are signatures of coherent coupling,
i.\,e. conversion of the absorbed $\omega_1$ into different
$\omega$, as depicted by squares. The mutual coupling within the
triplet (GX,\,XB,\,ET) is thus detected. The internal coupling in
QD1, i.\,e. between GX and XB, is expected\,\cite{KasprzakNPho11,
KasprzakNJP13}. Conversely, the transfer of coherence toward ET in
QD2 is more intriguing. Owing to a large frequency detuning between
ET and GX, radiative coupling\,\cite{MinkovPRB13} is excluded.
Coherent coupling between excitons in closely lying QDs stems from
their Coulomb interaction\,\cite{KasprzakNPho11, DanckwertsPRB06,
SpechtPRB15}, inducing biexciton shift or/and dipole-dipole
F\"{o}rster coupling. To distinguish between both mechanisms, in
Fig.\,\ref{fig:Fig3}\,b (top) we inspect the amplitude and phase of
the off-diagonal terms in the 2D FWM\,\cite{KasprzakNPho11}: the
former generates a double-peak and $2\pi$ spectral shift across the
off-diagonal signal, while the latter produces a single off-diagonal
peak with a $1\pi$ phase shift. This last scenario is indeed
revealed by the experimental data, indicating F\"{o}rster coupling
between QD1 and QD2. The spectral shift between the uncoupled and
coupled states is not detected owing to the large spectral detuning
of both QDs with respect to the dipole coupling strength estimated
to a few hundred of $\mu$eV. More involved multi-dimensional
spectroscopy\,\cite{SpechtPRB15} could provide additional insights
into the microscopic mechanism of the observed coupling. We note
that the broad operation bandwidth of the PT is here instrumental to
reveal such a largely off-resonant F\"{o}rster coupling between QD1
and QD2.

\begin{figure}[!ht]
\includegraphics[width=\columnwidth]{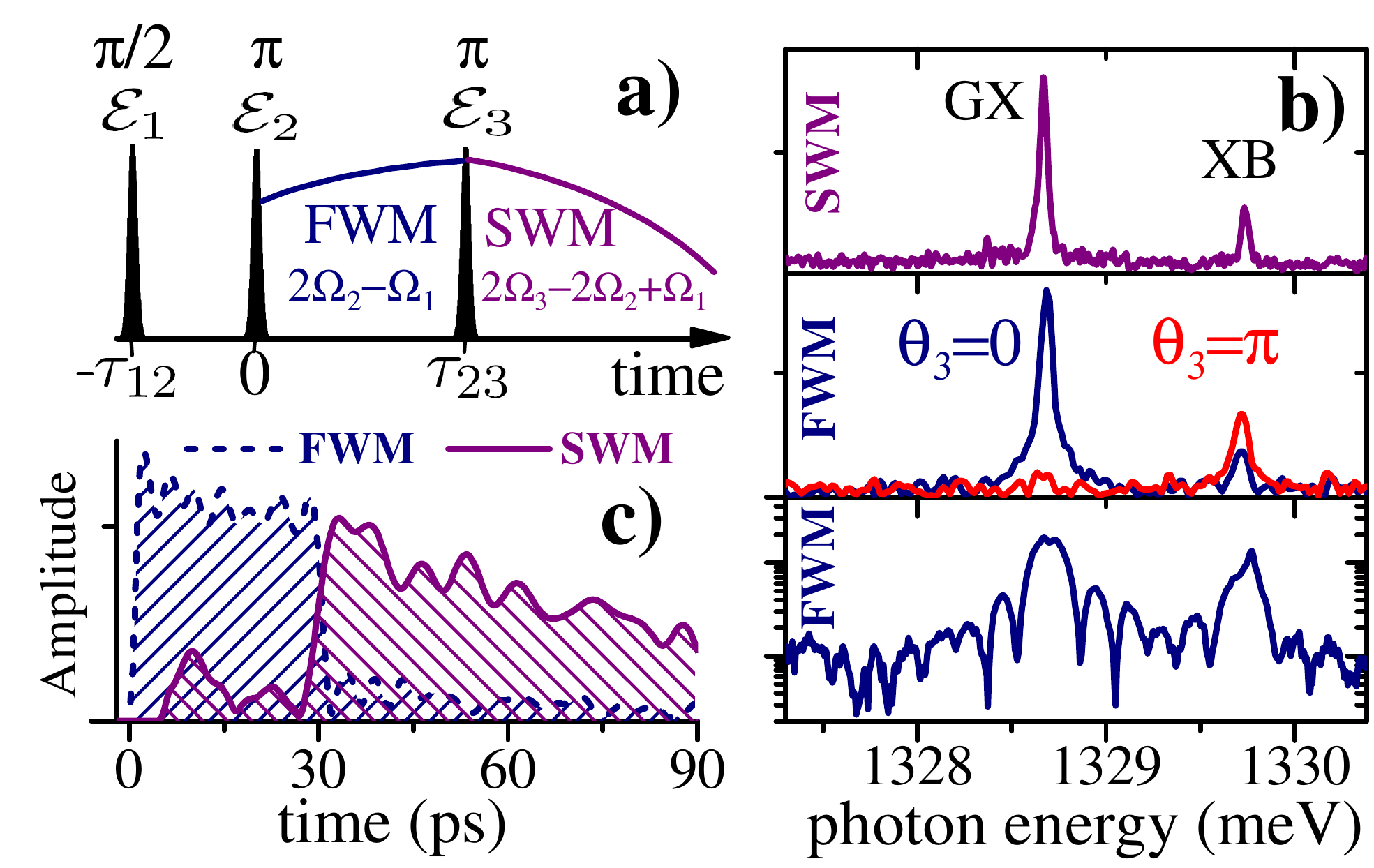}
\caption{\textbf{There-pulse coherent control of the
exciton-biexciton system in QD1.} a)\,A sketch of the pulse sequence
and their areas employed to drive the SWM and to manipulate the FWM
transient via FWM/SWM conversion scheme. b)\,Top:\,SWM spectrum of
GX and XB:\,$(\tau_{12},\,\tau_{23})=(15,\,30)\,$ps,
$(\theta_1,\,\theta_2,\,\theta_3)=(\pi/2,\,\pi,\,\pi)$.
Middle:\,Coherent control of the GX and XB by FWM/SWM switching:
$\theta_3=0$ and $\theta_3=\pi$ correspond to blue and red spectra,
respectively. Bottom:\,\ Manipulation of the FWM spectral response
of the GX and XB: $\theta_3=\pi$, $\tau_{23}=30\,$ps. c)\,Conversion
of the FWM to SWM observed on the GX in time
domain:\,$\theta_3=\pi$, $\tau_{23}=30\,$ps. \label{fig:Fig4}}
\end{figure}

In the following, we exploit a novel three-pulse
sequence\,\cite{Fras15} to demonstrate coherent control of the FWM
generated by the exciton-biexciton system of QD1. The protocol is
based on converting a desired amount of the FWM transient into the
six-wave mixing (SWM) one. The applied pulse sequence is depicted in
Fig.\,\ref{fig:Fig4}\,a. First, $\Ea$ and $\Eb$ drive degenerate
FWM, which is evolving during $\tau_{23}$. Then, $\Ec$ is used to
flip it into the specific SWM frequency. The amount of FWM converted
to SWM is governed by $\theta_3$. In particular, for $\theta_3=\pi$
the entirety of FWM is transferred to SWM. Such control via FWM/SWM
switching represents a step change with respect to past FWM
experiments, as it enables to accurately design the FWM response in
temporal and frequency domains, by tuning $\tau_{23}$ and
$\theta_3$. The SWM spectrum of the investigated exciton-biexciton
system driven at
$(\theta_1,\,\theta_2,\,\theta_3)=(\pi/2,\,\pi,\,\pi)$ is shown in
Fig.\,\ref{fig:Fig4}\,b (top). In the middle panel we present the
corresponding FWM: for $\theta_3=0$ and $\tau_{12}=3\,$ps (dark
blue) we recover the response as in Fig.\,\ref{fig:Fig1}\,b.
Instead, for $\theta_3=\pi$ (red) the FWM of the GX transition is
quenched, whereas the one of XB is enhanced. Such a FWM/SWM
switching at the GX transition is retrieved in the time domain in
Fig.\,\ref{fig:Fig4}\,c. At the arrival of $\Ec$ for
$\tau_{23}=30\,$ps the FWM is virtually suppressed, while the onset
of the SWM is observed. An abrupt cut of the signal in time domain
induces a particular FWM spectral line-shape\,\cite{Fras15} of GX
and XB, with a broadening of the main peaks and build up of
side-bands, exemplified in Fig.\ref{fig:Fig4}\,b (bottom) for
$\tau_{23}=30\,$ps. This effect, along with the temporal gating of
the FWM, is visualized in more details in the Supplementary Figure
\ref{fig:FigS4}, displaying FWM spectra and transients when varying
$\tau_{23}$. Note that the presented control scheme necessitates
addressing individual transitions in contrast to past SWM
experiments\,\cite{VossPRB02, TaharaPRL14, TaharaPRB14}.
Demonstration of SWM beyond two-level system, paves the way towards
three-dimensional spectroscopy (involving the interplay between
absorption, FWM and SWM) of quantum emitters, opening new frontiers
to monitor couplings and control the coherence in condensed matter.
Furthermore, prospective SWM studies on individual QDs could
elucidate their non-Markovian dephasing\,\cite{TaharaPRL14}.

To conclude, we performed coherent nonlinear spectroscopy of single,
strongly-confined excitons in InAs QDs embedded in PTs. Such
waveguide antennas offer a perfect interfacing of QD excitons with
external excitation, enabling to harvest their multi-wave mixing
responses with a sensitivity improved by up to four-orders of
magnitude with respect to a QD in bulk GaAs. Wave mixing
spectroscopy was employed not only to assess the coherence of
individual transitions, but also to ascertain inter-exciton
couplings with 2D FWM and to manipulate coherent response via
FWM/SWM switching. The broad operation bandwidth of the antenna
opens new possibilities for coherent spectroscopy of single
emitters, like performing multi-color multi-wave mixing. With such a
prospective Raman-type spectroscopy on single
QDs\,\cite{GammonScience97}, one could for example explore dynamics
of optical polarons\,\cite{HameauPRL99, ZibikPRB04} (coupled
exciton\,-\,optical phonon modes) and study propagation of acoustic
phonons between off-resonant, spatially distant excitons. The
broadband character of PTs also enables to perform wave-mixing
spectroscopy on excited states of a QD, to determine its level
structure and to retrieve related dephasings and coherent couplings.
Finally, by exploiting the energy tunability via
strain\,\cite{YeoNatNano14}, a pair of separate excitons could be
brought into the frequency resonance, so as to induce a radiative
coupling\,\cite{MinkovPRB13} mediated by the photons guided by the
wire antenna.

\emph{We acknowledge the support by the ERC Starting Grant PICSEN,
contract no.\,306387. Sample fabrication was carried out in the
``Plateforme Technologique Amont (PTA)" and CEA LETI MINATEC/DOPT
clean rooms. JK thanks Wolfgang Langbein for fruitful discussions
and continuous support.}

%------------------------------

\begin{center}
{\bf SUPPLEMENTARY MATERIAL}
\end{center}

\renewcommand{\figurename}{Supplementary Figure}
\renewcommand{\thefigure}{S\arabic{figure}}

\title{SUPPLEMENTARY MATERIAL\\ Harvesting, coupling and
control of single exciton coherences\\ in photonic waveguide
antennas}

\subsection{Sample fabrication.} The suspended PTs shown in Fig.\,\ref{fig:Fig1}\,a,
specifically developed for this work, offer improved mechanical
stability and robustness with respect to stand-alone ones\,$^{15}$.
They were fabricated from a planar GaAs sample containing a single
layer of InAs QDs grown by molecular beam epitaxy. After deposition
of a hard mask by e-beam lithography, metal deposition and lift-off,
the structures are defined by deep plasma etching with a controlled
under etching angle of $4^{\circ}$. The PTs investigated in the main
manuscript (Sample A) feature a length of 20$\,\mu$m and a top
diameter of 2.8$\,\mu$m. Around the QDs, the waveguide section
features a $0.25\,\mu$m diameter. Auxiliary pillars (used to reflect
the reference field, $\Er$) and PTs are respectively shifted by
$15\,\mu$m, and are organized in pattern of five-by-five PTs, with a
gradually varying PT top-diameters from 2$\,\mu$m to 4$\,\mu$m. By
performing photo-luminescence (not shown, acquired under pulsed,
non-resonant excitation) on the PT considered in the main
manuscript, we observe that the QD saturation is achieved for a pump
intensity as low as $0.4\mu$W. The transitions then feature a bright
emission, with spectrally-integrated count rate of $400\,$kHz.

\subsection{Coherence and FWM dynamics of InAs QDs in PTs: analysis of dephasing mechanisms. Supplementary examples.}

\begin{figure}[!ht]
\includegraphics[width=1.03\columnwidth]{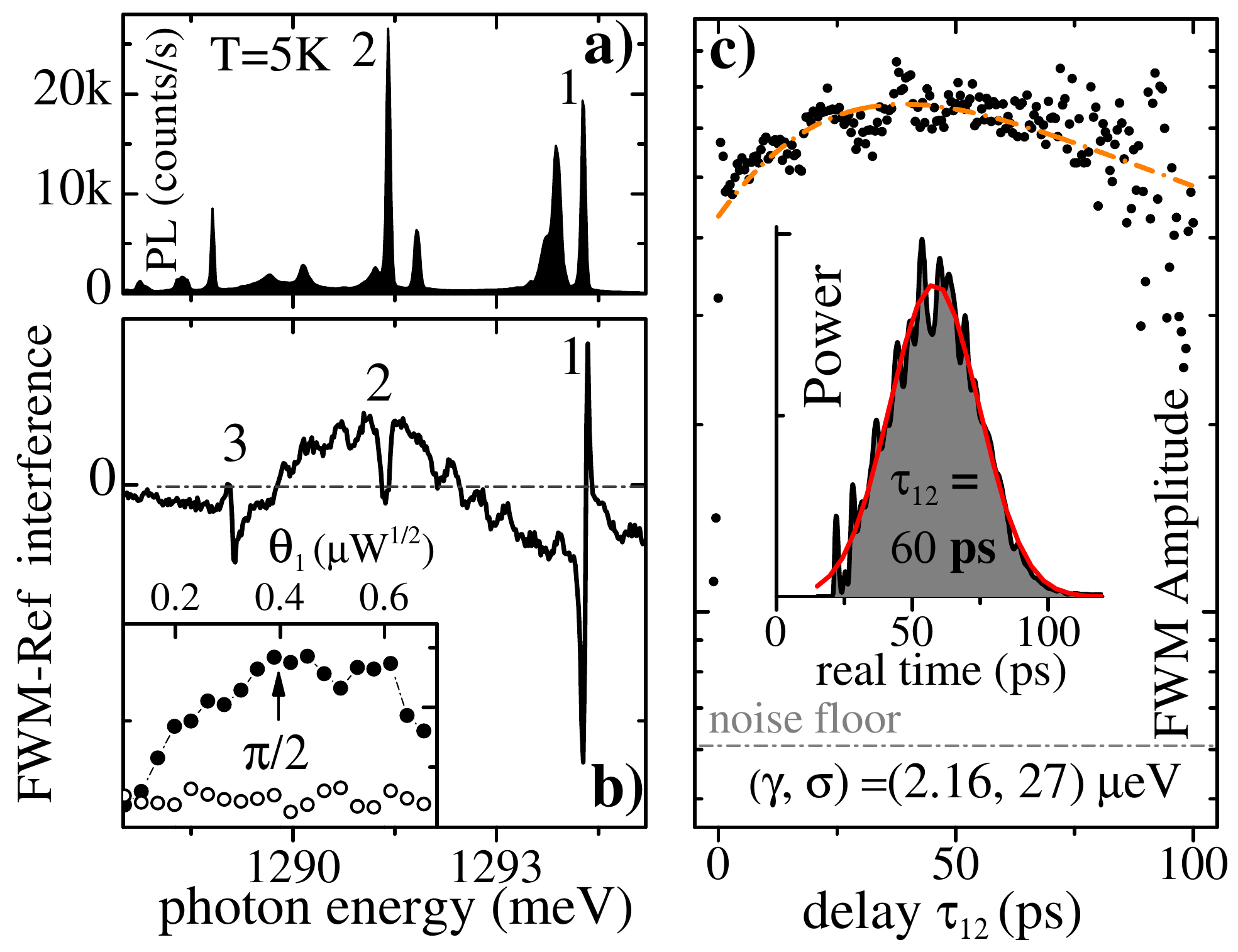}
\caption{\bf{PL and FWM spectroscopy of InAs QDs emitting at 960\,nm
embedded in a photonic trumpet. Sample B.}\label{fig:FigS1}}
\end{figure}

PTs enable to retrieve FWM from single InAs QDs within a broad
spectral range from 915\,nm to 960\,nm. To exemplify this broad-band
operation, in Fig.\,\ref{fig:FigS1} we present results obtained on
the sample (Sample B) containing QDs emitting at 960\,nm, with
respect to the one considered in the main manuscript (Sample A)
emitting around 930\,nm. In the Sample B, PTs have a length of
30$\,\mu$m, a top diameter of 4.5$\,\mu$m and a taper angle of
$\sim8^{\circ}$, and thus are bigger than in the Sample A.

In Fig.\,\ref{fig:FigS1}\,a we show a photoluminescence spectrum,
non-resonantly excited at 1.4 eV with intensity of 30\,nW, an order
of magnitude below the QD saturation. A spectral interference at the
FWM heterodyne frequency $2\Omega_2-\Omega_1$, obtained on the same
PT is shown in Fig.\,\ref{fig:FigS1}\,b. Three, spectrally sharp
features, correspond to the FWM from three individual transitions in
InAs QDs. It is worth to note that the PL and FWM spectra are not
the same. Transitions marked as $``1"$ and $``2"$ are observed in
both experiments. Conversely, $``3"$, detected in FWM, does not have
its counterpart in the PL. FWM amplitude of the transition $``1"$ as
a function of $\Ea$ pulse area $\theta_1$ is shown in the inset. FWM
reaches its maximum, corresponding to the $\theta_1=\pi/2$ area, for
$\Ea$ intensity of only $0.2\,\mu$W, indicating an excellent
coupling between the $\Eo$ driving fields and the exciton
transition. Further increase of $\theta_1$ results in the Rabi
flopping and decrease of the FWM amplitude.

\begin{figure}[!ht]
\includegraphics[width=1.02\columnwidth]{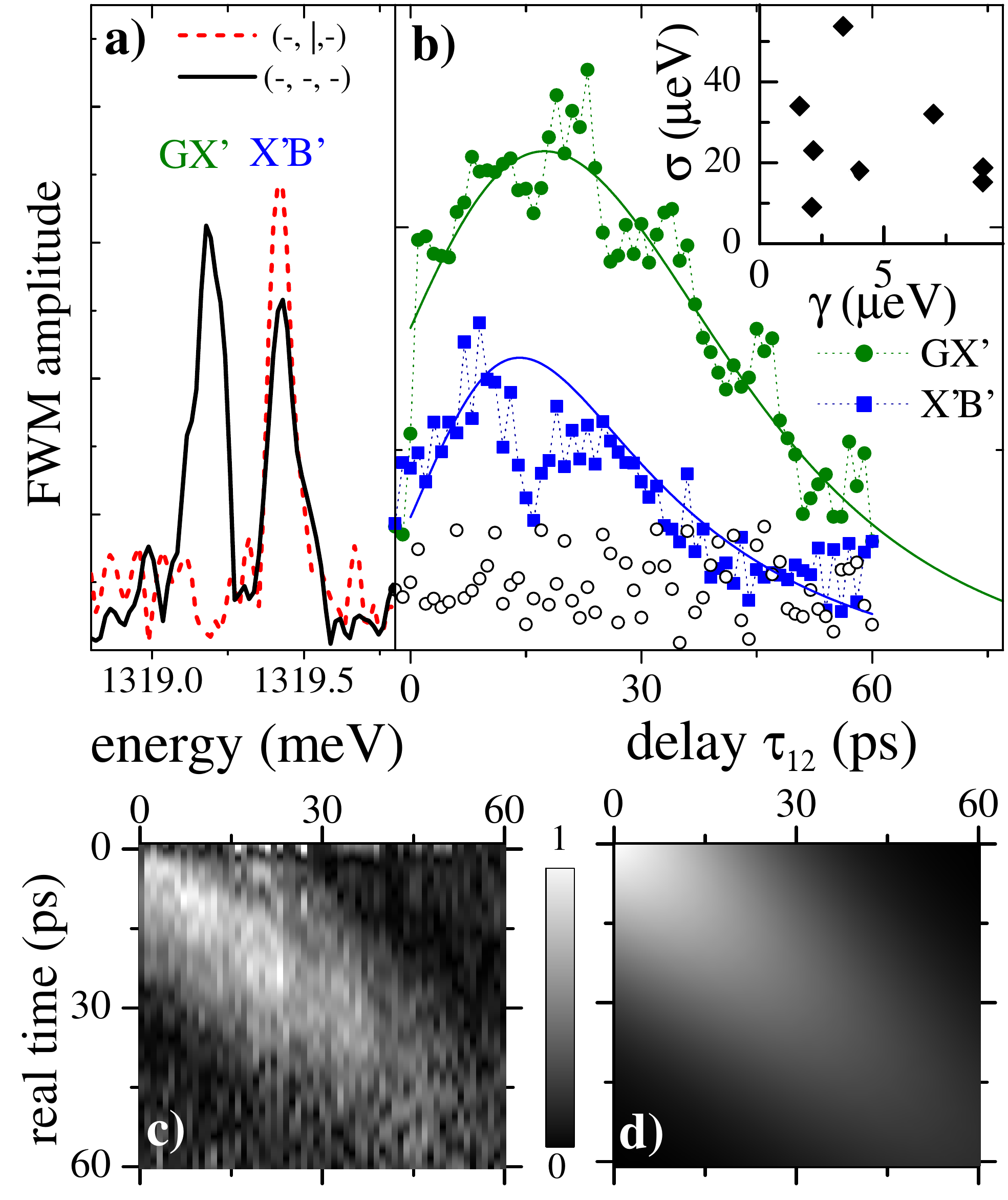}
\caption{{\bf Coherence and FWM dynamics measured on an
exciton-biexciton system showing an increased inhomogeneous
broadening $\sigma$ with respect to Fig.\,\ref{fig:Fig2} and
Fig.\,\ref{fig:FigS1}.}\label{fig:FigS2}}
\end{figure}

In order to infer dephasing mechanisms, we have first examined the
time-resolved FWM for fixed $\tau_{12}=60\,$ps, displayed as an
inset in Fig.\,\ref{fig:FigS1}\,c. The response clearly reveals a
Gaussian photon echo (the red trace corresponds to a Gaussian fit),
with the temporal FWHM spread of t$_{\rm inh}=56\,$ps for the FWM
amplitude. Such an echo in the FWM transient is a fingerprint of the
spectral inhomogeneous broadening $\sigma$. Applying the model
presented in Ref.\,[20], it is evaluated as $\sigma=\hbar\sqrt{8{\rm
ln}2}/{\rm t}_{\rm inh}=27\,\mu$eV. On a single exciton level,
$\sigma$ is attributed to the spectral wandering, occurring within
the integration time\,$^{20}$. By varying the latter from 1\,ms to
1\,s we observe virtually the same echo behavior. We thus conclude
that the spectral fluctuations of this transition occur at the
sub-ms timescale. Spectral positions of the transition ``1"
fluctuate with a characteristic spread of $\sigma$ over the
measurement. Despite their temporal separation, they all interfere
with the reference field $\Er$ and give rise to the involved
spectral shape of the resulting time-averaged interferogram. After
applying Fourier-transform in the spectral interferometry algorithm,
this yields a Gaussian response in time centered at t$=\tau_{12}$,
i.\,e. photon echo. We note that excitons in all investigated PTs
(statistics of $\gamma$ and $\sigma$ is shown in the inset of
Fig.\,\ref{fig:FigS2}\,b) displayed a measurable echo in their
coherence dynamics, yielding $\sigma$ up to $50\,\mu$eV.

To assess the coherence dynamics, we have measured $\tau_{12}$ delay
dependence of the FWM amplitude, displayed in
Fig.\,\ref{fig:FigS1}\,c. Suppression of the FWM for negative delays
indicates lack of two-particle, biexcitonic state. We thus attribute
the investigated transition $``1"$ to a charged exciton. In the FWM
delay dynamics, we observe initial rise, which is due to the echo
formation, followed by an exponential decay. To simulate coherence
evolution (depicted with the orange dashed line, also see
Fig.\,\ref{fig:Fig2}\,a and Fig.\,\ref{fig:FigS2}\,b) of such
inhomogeneously broadened transition, we implement the model
presented in Ref.\,[20]. Bearing in mind previously determined
$\sigma$, we retrieve the dephasing time T$_2=(610\,\pm\,20)\,$ps as
the only fitting parameter, corresponding to the homogeneous
broadening of $\gamma=(2.16\,\pm\,0.07)\mu$eV.

In Fig.\,\ref{fig:FigS2}\,a we present FWM spectra in
$(\Ea,\,\Eb,\,\Er)=(\polho,\,\polho,\,\polho)$ (black solid) and
$(\polho,\,\polve,\,\polho)$ (red dashed) configuration on another
PT in the Sample A, having a top-diameter of $2.9\,\mu$m. Using the
FWM polarization selection rules (see Fig.\,\ref{fig:FigS3}), a pair
of transitions (GX$'$,\,X$'$B$'$) is identified as an
exciton-biexciton system, with a renormalization energy of
$\Delta'=-0.24\,$meV. The retrieved coherence dynamics is shown in
Fig.\,\ref{fig:FigS2}\,b, yielding in this case, $(\gamma_{\rm GX'},
\sigma_{\rm GX'})=(7,\,33)\,\mu$eV and $(\gamma_{\rm X'B'},
\sigma_{\rm X'B'})=(7,\,58)\,\mu$eV. Clearly, larger $\sigma$
deteriorates the coherence of both transitions, resulting in larger
$\gamma$ with respect to the example investigated in the main
manuscript (Fig.\,\ref{fig:Fig2}). Note, a larger inhomogeneous
broadening of X$'$B$'$ with respect to the GX$'$, indicating
anti-correlation in spectral wandering of the exciton and the
biexciton level. In the coherence dynamics we also identify
exciton-biexciton beating with a period of
$|2\pi\hbar/\Delta'|=17\,$ps. The noise floor is given by open
circles.

The time-resolved FWM amplitude of the GX$'$ as a function of
$\tau_{12}$ is shown in Fig.\,\ref{fig:FigS2}\,c. Due to a large
$\sigma$ (and thus sufficiently narrow width of the echo in time)
such a map explicitly shows formation of the photon echo\,$^{20}$.
Namely, we see the shift of the FWM maximum in real time t according
to $\tau_{12}$, forming a Gaussian centered along the diagonal line
$\tau_{12}=$t, as reproduced by the simulation shown in
Fig.\,\ref{fig:FigS2}\,d.

\subsection{Identification of the exciton-biexciton system via FWM selection rules}

\begin{figure}[!ht]
\includegraphics[width=0.9\columnwidth]{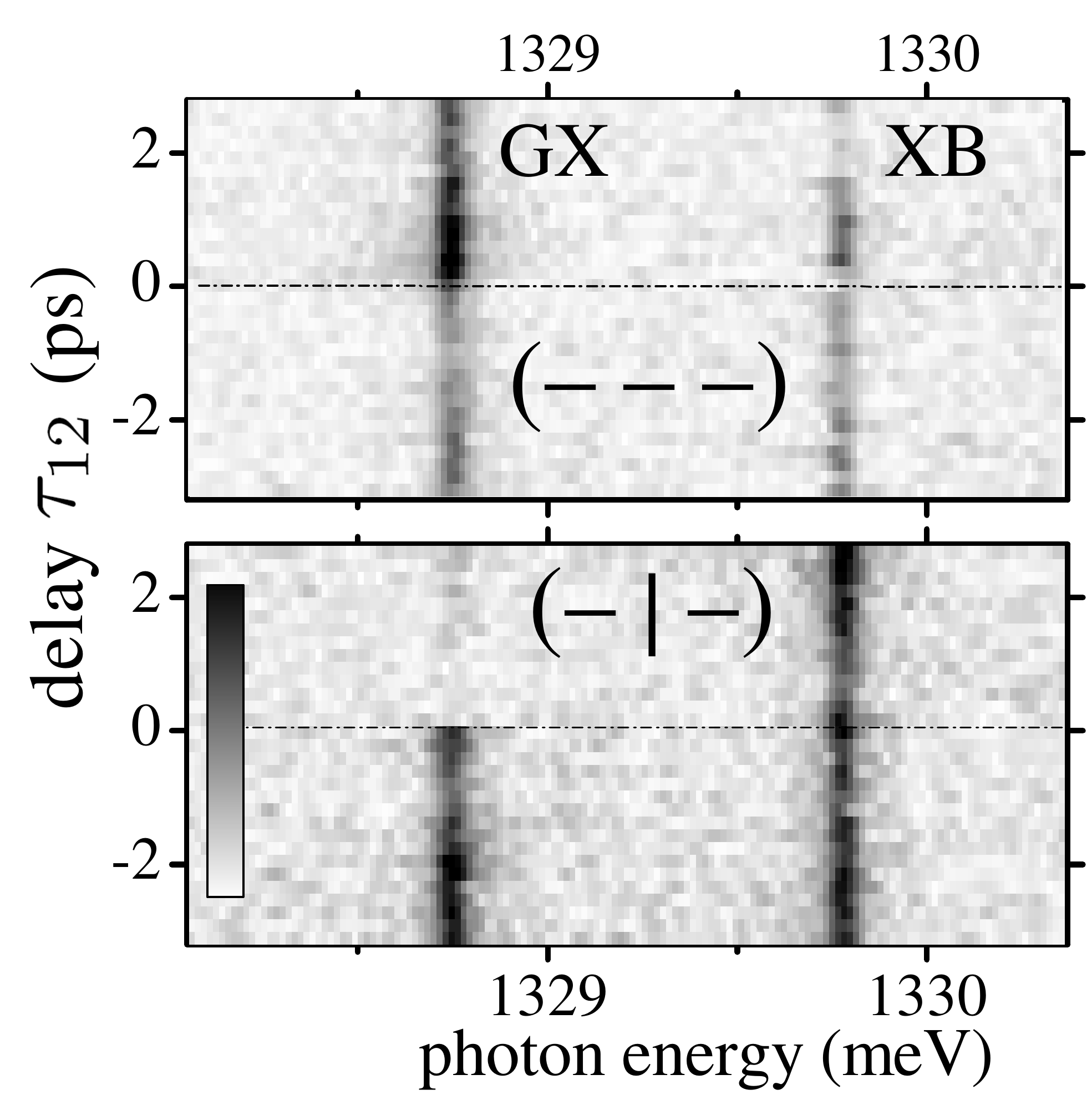}
\caption{{\bf Identification of GX and XB as a ground state-exciton
and exciton-biexciton transiton in a single InAs QD.} Delay
dependence of the FWM for -3.2\,ps\,$<\tau_{12}<\,2.8\,$ps, for
co-linearly polarized $\Ea$, $\Eb$ and the reference
$(\polho,\,\polho,\,\polho)$ (top) and for cross-polarized $\Eb$;
$(\polho,\,\polve,\,\polho)$ (bottom). Detection along ($\polho$)
direction.\label{fig:FigS3}}
\end{figure}

The resonances discussed in the main manuscript at 1328.74\,meV and
1329.8\,meV are recognized as ground state\,-\,exciton (GX) and
exciton\,-\,biexciton (XB) transitions. We apply the FWM
polarization and delay selection rules presented in Ref.\,[20]. In
Fig.\,\ref{fig:FigS3}, we show FWM amplitude of GX and XB, as a
function of $\tau_{12}$, around zero delay for co- and
cross-linearly polarized $\Ea$ and $\Eb$, for
$(\theta_1,\,\theta_2)=(\pi/5,\,2\pi/5)$. For negative delays, the
FWM is the same for both polarization configurations: $\Eb$ arrives
first and induces a two-photon coherence (TPC) between G and B.
$\Ea$ converts TPC into the FWM released with equal amplitudes on GX
and XB. Conversely, for positive delays the FWM can be generated in
two manners. For co-linearly polarized $\Ea$ and $\Eb$ the FWM is
created via density grating, with the amplitude twice larger at GX
than at XB (in the $\chi^3$, perturbative regime). Instead, for
cross-polarized excitation, the signal is created via Raman
coherence between GX and GY, such that the signal is released
uniquely at XB. Finally, we note that we also measured the
fine-structure beating with a period of 50\,ps, yielding the
splitting between GX and GY of $\delta=83\,\mu$eV (not shown).

\subsection{Temporal gating and engineering of the FWM spectral lineshape via FWM/SWM switching}

\begin{figure}[!ht]
\includegraphics[width=1.05\columnwidth]{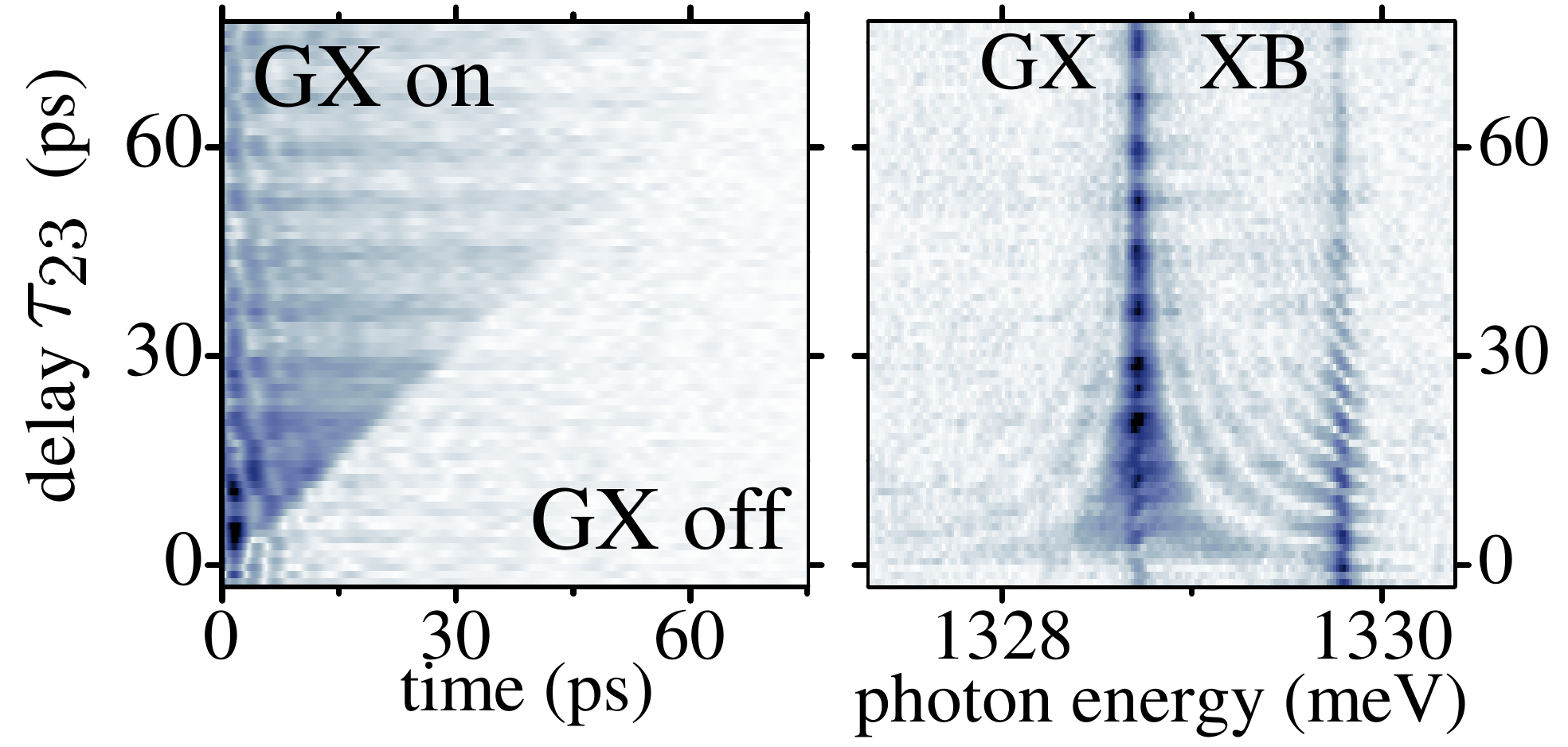}
\caption{{\bf Coherent control of the FWM signal of the
exciton-biexciton system in QD1 (Sample A), via FWM/SWM switching
mechanism.} \label{fig:FigS4}}
\end{figure}

FWM amplitude of the GX transition in QD1, discussed in the main
manuscript, as a function of the real time and delay $\tau_{23}$ of
the control pulse $\Ec$ is shown in the left panel in
Fig.\,\ref{fig:FigS4}. FWM is present uniquely prior to the arrival
of $\Ec$, showing temporal gating of the signal. Linear color scale.
In the right panel we show FWM spectral amplitudes. Owing to an
abrupt suppression of the FWM in the real time, a considerable
spectral broadening and sidebands for both GX and XB are observed
for initial delays $\tau_{23}$, followed by a gradual recovery of
the Lorentzian lineshapes with increasing $\tau_{23}$.

\end{document}